\documentclass[letterpaper]{article}
\usepackage{aaai20}
\usepackage{times}
\usepackage{helvet}
\usepackage{courier}
\usepackage[hyphens]{url} 
\usepackage{graphicx}
\usepackage{graphicx}
\usepackage{array}
\usepackage{balance}
\usepackage{subfigure}
\usepackage{multirow}
\usepackage{float}
\usepackage{color}
\usepackage{soul}
\usepackage{mdframed}
\usepackage{amsopn}
\usepackage{mathrsfs}
\usepackage{mathtools}
\usepackage{amsmath}
\usepackage{amssymb}
\usepackage{bbm}
\usepackage{arydshln}
\usepackage{hyperref}
\usepackage{multicol}
\usepackage{blkarray}
\usepackage{enumerate}
\usepackage{courier}
\usepackage{rotating}
\usepackage{booktabs}
\usepackage{diagbox}
\usepackage{fancybox}
\usepackage{minibox}
\usepackage{cases}
\usepackage[noend]{algpseudocode}
\usepackage{algorithm}

\newcommand{\citet}[1]{\citeauthor{#1} \shortcite{#1}}

\newtheorem{definition}{Definition}

\newcommand{\approach}{{NSER}}

\title{Neural-Symbolic Reasoning over Knowledge Graph for \\Multi-stage Explainable Recommendation}
\author{Yikun Xian,  \textsuperscript{\rm}
Zuohui Fu,  \textsuperscript{\rm}
Qiaoying Huang \textsuperscript{\rm}
S. Muthukrishnan \textsuperscript{\rm}
Yongfeng Zhang \textsuperscript{\rm}\\
\textsuperscript{\rm}Department of Computer Science\\ Rutgers University, New Brunswick, NJ, USA\\
siriusxyk@gmail.com, zuohui.fu@rutgers.edu, qh55@cs.rutgers.edu\\ muthu@cs.rutgers.edu, yongfeng.zhang@rutgers.edu \\
}

\begin{document}

\maketitle

\begin{abstract}
Recent work on recommender systems has considered external knowledge graphs as valuable sources of information, not only to produce better recommendations but also to provide explanations of why the recommended items were chosen. Pure rule-based symbolic methods provide a transparent reasoning process over knowledge graph but lack generalization ability to unseen examples, while deep learning models enhance powerful feature representation ability but are hard to interpret. Moreover, direct reasoning over large-scale knowledge graph can be costly due to the huge search space of pathfinding. We approach the problem through a novel coarse-to-fine neural symbolic reasoning method called \approach{}. It first generates a coarse-grained explanation to capture abstract user behavioral pattern, followed by a fined-grained explanation accompanying with explicit reasoning paths and recommendations inferred from knowledge graph. We extensively experiment on four real-world datasets and observe substantial gains of recommendation performance compared with state-of-the-art methods as well as more diversified explanations in different granularity. 
\end{abstract}

\section{Introduction}

Explainable recommendation has attracted increasing attention from both academic and industry communities \cite{zhang2018explainable}, 
which highlights the importance of system transparency and targets to provide  more informed recommendation results and satisfactory user experience \cite{liao2018tscset}.
In this regard, knowledge graphs (KG) have recently come to prominence for building explainable recommender system, as the graph structure empowers the ability to trace reasoning paths behind recommendations. 
Early works \cite{catherine2017explainable,catherine2016personalized} propose to model user behavior by a set of symbolic rules from knowledge graph. 
Despite their transparency, these methods heavily depend on the handcrafted rules, which makes them difficult to generalize to unseen behavior correlations. 
With the advances of deep learning \cite{fu2020absent}, several neural-based methods have been proposed to incorporate knowledge graphs into recommender systems, 
in either \textbf{\textit{pre-defined}} \cite{ma2019jointly,wang2018explainable} or \textbf{\textit{post-hoc}} \cite{ai2018learning} manner. 
However, they fail to explicitly model the path routing process, which impedes the transparency of the recommendations. 
Recent work \cite{xian2019kgrl} study \textbf{\textit{in-progress}} explainable recommendation by adopting reinforcement learning for path inference in knowledge graph,
but the recommendation performance is sacrificed by the explanation generation process and the paths are not sufficiently diversified.

\begin{figure}
    \centering
    \includegraphics[width=0.85\linewidth]{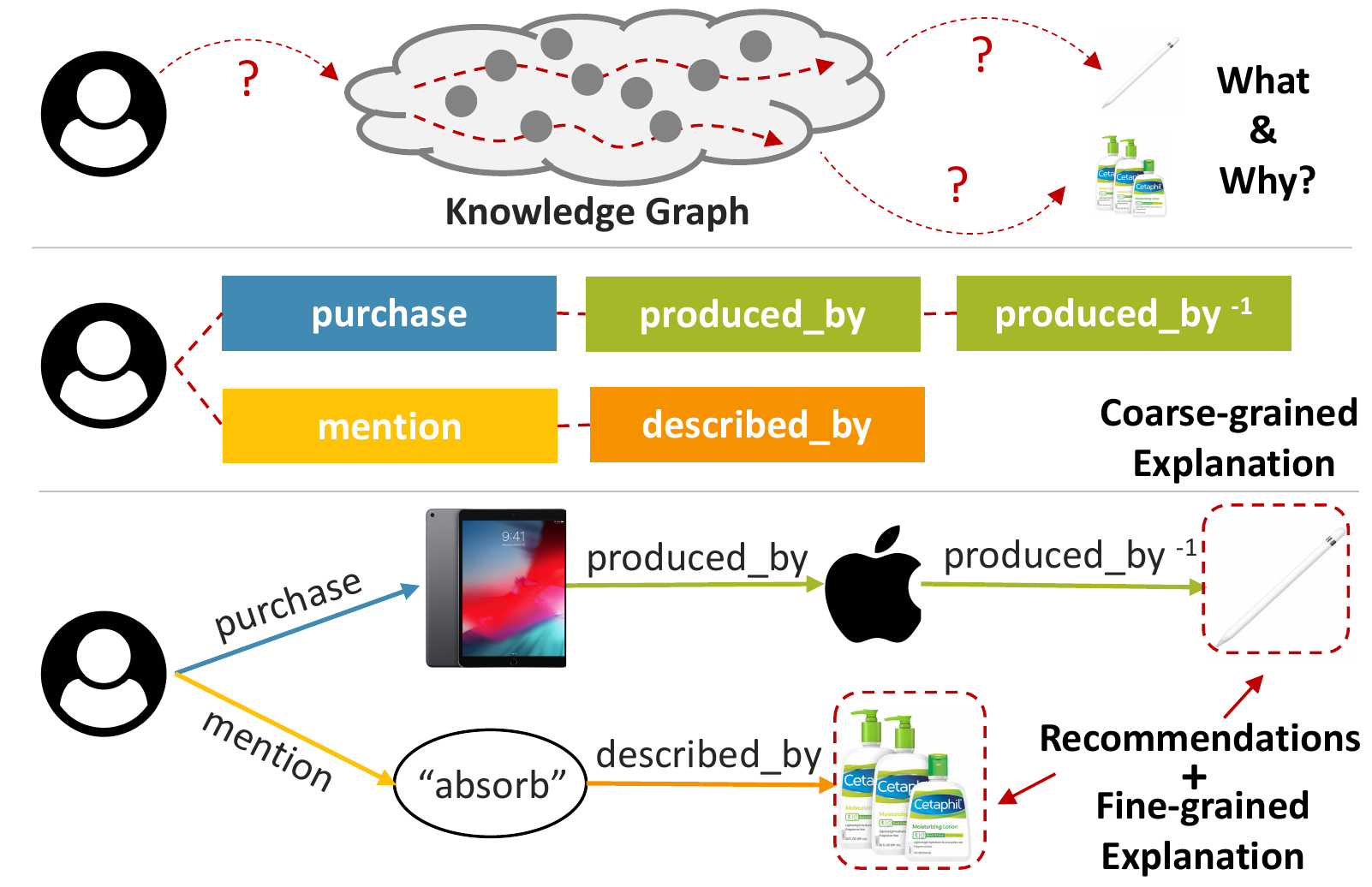}
    \vspace{-8pt}
    \caption{A coarse-to-fine process of neural-symbolic reasoning over knowledge graph for explainable recommendation. The coarse-grained explanation provides abstract and representative user patterns, which is then invoked to produce fine-grained reasoning paths for recommendations.}
    \label{fig:motivation}
    \vspace{-10pt}
\end{figure}

Other previous work has exploited symbolic rules using neural networks to encode prior knowledge to advance the intent representation and performance with good interpretability. %
\cite{karpathy2015visualizing} empirically expose interactions within neural network structures. \cite{santa2018neural} develop neural network modules to compose boolean algebraic operations on visual classifiers, with an ability to synthesize them for complex visual concept expressions. However, despite success in visual reasoning tasks, no existing work seeks to exploit these neural symbolic methods in explainable recommendations.

To leverage the best of both worlds, we propose a coarse-to-fine neural symbolic reasoning method called \approach{} to infuse explicit symbolic reasoning into the neural modeling process for explainable recommendation, which induces an intermediate reasoning template before concrete path reasoning over knowledge graph for recommendation and explanation.
As illustrated in Figure \ref{fig:motivation}, \approach{} first learns to generate a coarse-grained explanation that depicts abstract but representative user behavioral pattern.
Such explanation is characterized by composition of neural relation modules, which facilitates personalization of both explanation and recommendation by refactoring these modules.
Then, \approach{} produces a fine-grained explanation containing concrete reasoning paths based on the user pattern, and the end items in the paths are naturally acquired for recommendation.
The use of such two-stage framework not only yields high-quality recommendation results, but also provides a form of hierarchical explanations for better interpretability.

The contributions of this paper are threefold.
First, we emphasize the importance of explainable recommendation with knowledge graph, especially in-progress explanations that fully match the actual decision making for recommender systems, as opposed to predefined or post-hoc explanations.
Second, we marry the merits of deep learning with the interpretability of symbolic graph reasoning and propose a two-stage coarse-to-fine reasoning model for the problem.
Third, we experiment on four real-word benchmarks showing that our model yields both good recommendation results and versatile hierarchical explanations.
To the best of our knowledge, this is the first paper that studies the generation of coarse-to-fine explanations for recommendations via multi-stage neural symbolic reasoning process.

\section{Problem Formulation}

\begin{figure*}[t]
    \centering
    \includegraphics[width=0.95\textwidth]{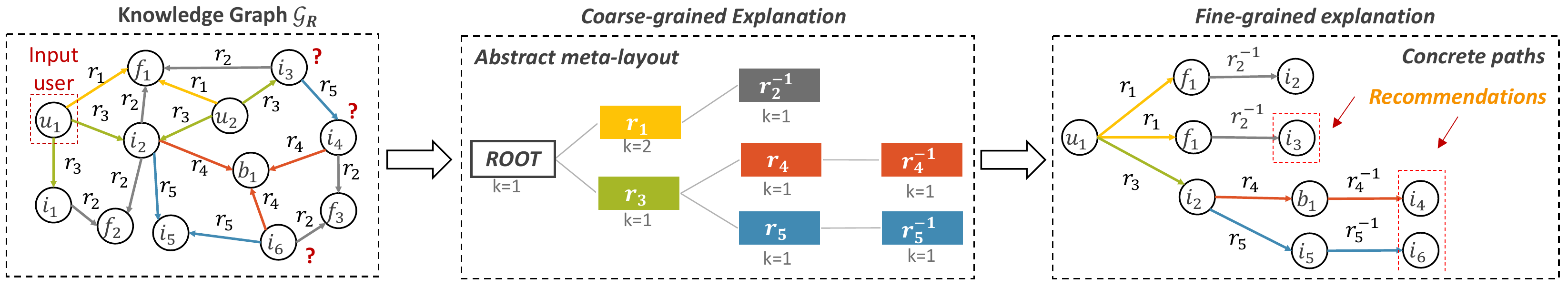}
    \vspace{-15pt}
    \caption{The two-stage pipeline of our method to generate coarse-to-fine explanations and recommendations. Symbols in the knowledge graph are represented as entity embeddings and neural relation modules. A coarse-grained explanation (abstract meta-layout) is first generated to describe representative user patterns for recommendation. A following fine-grained explanation is produced based on the pattern to provide explicit reasoning paths whose end nodes serve as recommendations.}
    \label{fig:method}
    \vspace{-5pt}
\end{figure*}

A \textbf{recommendation knowledge graph} $\mathcal{G}_\mathrm{R}$ is defined as a set of triples, $\mathcal{G}_\mathrm{R} = \{(e_h,r,e_t)\mid e_h,e_t\in\mathcal{E},r\in\mathcal{R}\}$ with entity set $\mathcal{E}$ and relation set $\mathcal{R}$.
Each triple $(e_\mathrm{h},r,e_\mathrm{t})$ represents a fact indicating head entity $e_\mathrm{h}$ interacts with tail entity $e_\mathrm{t}$ through relation $r$.
The entity set $\mathcal{E}$ contains $N_\mathcal{E}$ disjoint subsets of entities, i.e., $\mathcal{E} = \mathcal{U} \cup \mathcal{I} \cup \mathcal{E}_3 \cup\cdots\cup \mathcal{E}_{N_\mathcal{E}}$, where $\mathcal{U}$ and $\mathcal{I}$ respectively stand for ``users'' and ``items'' in the context of recommendation.
We are particularly interested in the relation of user--item interaction, denoted by $r_{ui}\in\mathcal{R}$, which in practice represents ``purchase'' in e-commerce recommendation.
Formally, the problem of explainable recommendation over knowledge graph is defined as follows.
\begin{definition}{(Problem Definition)}
Given an incomplete knowledge graph $\mathcal{G}_\mathrm{R}$, the goal is to recover all missing facts $\{(u,r_{ui},i) \mid u\in\mathcal{U},i\in\mathcal{I}\}$ such that each pair $(u,i)$ is derived from a reasoning path from $u$ to $i$, which serves as an explanation towards the recommendation $i$ for user $u$.
\end{definition}

The challenges of the problem are threefold.
First, the explainable path $L_{ui}^\pi$ is expected to be generated in an \textbf{in-progress} manner and is supposed to explicitly explain how the model makes decisions step by step over the knowledge graph towards a potentially good item for recommendation.
Second, the search space to generate explainable paths can be very large in large-scale knowledge graphs. It is necessary to narrow down the scope through a coarse-to-fine process that first generate abstract patterns to sketch user behavior prior to reasoning about concrete paths for recommendation.
Third, the method should support personalization in both recommendation and explanation.

\section{Methodology}
We propose a method called Neural-Symbolic Explainable Recommendation (\approach) that takes advantage of both interpretable symbolic reasoning as well as of powerful neural representation learning.
As shown in Figure \ref{fig:method}, our model makes in-progress explainable recommendations using a two-stage coarse-to-fine reasoning process over KG. 
It first learns to generate coarse-grained patterns by composing various reasoning components of relation symbols, and then reasons about fine-grained paths consisting of entity symbols for recommendation.
Both pattern and path are treated as explanations, where the former represents abstract user behaviors while the latter describes concrete decisions made by the system.
In the following, we first introduce neural representation learning for symbols in KG, followed by the two-stage process for explainable recommendation.

\subsection{Neural-Symbolic Representation Learning}
In the knowledge graph, symbols such as entities and relations can directly be used for reasoning via predefined rules, but such rule-based method lack any generalization ability to unseen rules.
Therefore, we neuralize these symbols via deep representation learning and make these neural symbols serve as building blocks in the subsequent reasoning stage.

To be specific, each entity $e\in\mathcal{E}$ is regarded as a $d$-dimensional vector, denoted by $\mathbf{e}\in\mathbb{R}^d$, known as an \textbf{entity embedding}.
In the task of path reasoning for recommendation, we assume that the next entity $e'$ leading to a good target item is determined by relation $r$ given the start user $u$ and the preceding entity $e$.
Hence, we treat each relation $r\in\mathcal{R}$ as a reasoning component or a function, denoted by $\phi_r:\mathbb{R}^d\times\mathbb{R}^d\longmapsto\mathbb{R}^d$, which maps vectorized user and the preceding entity $(\mathbf{u}, \mathbf{e})$ to the successor entity embedding $\mathbf{e}'$.
In this work, we approximate $\phi_r$ by a neural network parameterized by $\Theta_r$, written as $\mathbf{e}'=\phi_r(\mathbf{u}, \mathbf{e};\Theta_r)$, and call it the \textbf{neural relation module} of relation $r$.
We treat the linear composition of such modules as an instantiation of the corresponding metapath. A tree-structured composition of modules constructs coarse-grained abstract pattern, which will later be invoked to generate fine-grained reasoning paths for explainable recommendation. %

Let $\Theta_{\mathcal{G}_\mathrm{R}}=\{\mathbf{e}|\forall e\in \mathcal{E}\}\cup\{\Theta_r|\forall r\in\mathcal{R}\}$, the set of all learnable parameters for the neural symbolic representation. The goal of neural-symbolic representation learning is to optimize over $\Theta_{\mathcal{G}_\mathrm{R}}$ such that neural symbols can be composed in a certain way leading to good reasoning ability for decision making.
In particular, let $\mathcal{I}_u^+$ be the set of positive items of user $u$, and $\mathcal{L}_u^\pi=\{L_{ui}^\pi|i\in\mathcal{I}_u^+\}$, the set of user--item paths from $u$ to any positive items via metapath $\pi$.
Now, given a user $u$ and metapath $\pi$, we aim to minimize the negative log likelihood of all positive paths being generated, with the following objective function $\ell_\mathrm{path}(u,\pi)$:
\begin{align}
\footnotesize
\hspace{-3pt}\ell_\mathrm{path}(u,\pi)  
 &= \underset{L_{ui}^\pi\sim\mathcal{L}_{u}^\pi}{\mathbbm{E}}\left[ -\sum_{j=1}^{|\pi|}\log P\left(e_j \mid u,e_{j-1},r_j\right) \right] \label{eq:entity2}
\end{align}
With the help of the neuralized entity embeddings and neural relation modules, the probability term  in Eq.~\ref{eq:entity2} can be approximated as 
$P\left(e_j \mid u,e_{j-1},r_j\right) \approx  \frac{\exp(s(\mathbf{e}_j))}{\sum\limits_{e\in \mathcal{E}_j}\exp(s(\mathbf{e}))}$,
where $s(\cdot) = \left\langle \cdot, \phi_{r_j}(\mathbf{u},\hat{\mathbf{e}}_{j-1};\Theta_{r_j}) \right\rangle$
and $\mathcal{E}_j\subseteq\mathcal{E}$ denotes the subset of all entities that share the same type as $e_j$.

Optimizing over the loss function in Eq.~\ref{eq:entity2} empowers the neural relation modules to fit positive paths, but it fails to distinguish the relative importance among them.
Thus, we impose an additional ranking loss on the final output entity.
Suppose there is an off-the-shelf but less accurate teacher model $h$ that estimates the similarity between a user $u$ and an item $i$, i.e., $h:(u,i)\mapsto[0,1]$.
We define the negative item set $\mathcal{I}_{ui}^-$ with respect to an item $i$ and $\mathcal{I}_u^+$ to be $\mathcal{I}_{ui}^-=\{i^- \mid h(u,i)>h(u,i^-),i^-\in\mathcal{I}_u^+\}$.
The ranking loss $\ell_\mathrm{rk}(u,\pi)$ with respect to user $u$ and metapath $\pi$ is defined as:
\begin{equation}\label{eq:loss_rank}
\footnotesize
\hspace{-3pt}
\ell_\mathrm{rk}(u,\pi) = \underset{L_{ui}^\pi\sim\mathcal{L}_u^\pi}{\mathbbm{E}}\left[ \underset{i^-\sim \mathcal{I}_{ui}^-}{\mathbb{E}} \left[ \sigma\left( \left\langle\hat{\mathbf{e}}_{|\pi|},\mathbf{i}^-\right\rangle - \left\langle\hat{\mathbf{e}}_{|\pi|},\mathbf{i}\right\rangle \right)\right] \right],
\end{equation}
where $\sigma(\cdot)$ is sigmoid function, and $\hat{\mathbf{e}}_{|\pi|}$ is the output entity embedding from the last neural relation module $\phi_{r_{|\pi|}}$.

By aggregating Eq.~\ref{eq:entity2} and \ref{eq:loss_rank} across all users and metapaths in the knowledge graph $\mathcal{G}_\mathrm{R}$, the overall objective becomes:
\begin{equation}\label{eq:loss_all}
\ell_\mathrm{all}\left(\Theta_{\mathcal{G}_\mathrm{R}}\right) = \sum_{u\in\mathcal{U},\pi\in\mathcal{G}_\mathrm{R}} \ell_\mathrm{path}(u,\pi)+\lambda \ell_\mathrm{rk}(u,\pi),
\end{equation}
where $\lambda$ is the weighting factor for the ranking loss.

\subsection{Neural-Symbolic Explainable Recommendation}
\begin{algorithm}[t]
\caption{Abstract Meta-Layout Generation}
\label{alg:program-gen}
\begin{algorithmic}[1]
\footnotesize
\State \textbf{Input:} metapaths $\{\pi_1,\ldots,\pi_M\}$ with $\{y_1,\ldots,y_M\}$
\State \textbf{Output:} abstract meta-layout $T$
\State Initialize $T$ by merging $\{\pi_1,\ldots,\pi_M\}$
\State $\forall x\in \text{Leaves(T)}, \pi_j$: \textbf{if} $x\in \pi_j$ \textbf{then} $k_x\gets y_j$
\State \Call{RecursiveUpdate}{$T.root$}
\State \Return $T$
\vspace{0.2em}
\Procedure{RecursiveUpdate}{node $x$}
  \State $\forall c\in \text{Children}(x)$, \Call{RecursiveUpdate}{$c$}
  \If{$x$ is root} $k_x\gets 1$ \Return \EndIf
  \State $C\gets \{k_c \mid c\in \text{Children}(x), k_c>0\}$
  \If{$C = \emptyset$} $k_x \gets 0$
  \Else~$k_x \gets \min(C)$, $\forall c\in \text{Children}(x): k_c \gets \lfloor k_c/k_x \rfloor$
  \EndIf
\EndProcedure
\end{algorithmic}
\end{algorithm}
\setlength{\textfloatsep}{8pt}%

\begin{table*}[t]
\scriptsize
\begin{tabular*}{\textwidth}{c|p{2.3em}p{2.3em}p{2.3em}p{2.3em}|p{2.3em}p{2.3em}p{2.3em}p{2.3em}|p{2.3em}p{2.3em}p{2.3em}p{2.3em}|p{2.3em}p{2.3em}p{2.3em}p{2.3em}}
\hline
Dataset & \multicolumn{4}{c|}{\textbf{CDs \& Vinyl}} & \multicolumn{4}{c|}{\textbf{Clothing}} & \multicolumn{4}{c|}{\textbf{Cell Phones}} & \multicolumn{4}{c}{\textbf{Beauty}} \\
\hline 
Measures ($\%$) & NDCG  & Recall & HR    & Prec. & NDCG  & Recall & HR   & Prec. & NDCG  & Recall & HR    & Prec. & NDCG  & Recall & HR     & Prec. \\
\hline 
DeepCoNN       & 4.218 & 6.001 & 13.857 & 1.681 & 1.310 & 2.332 & 3.286 & 0.229 & 3.636 & 6.353 & 9.913  & 0.999 & 3.359 & 5.429  & 9.807  & 1.200\\
CKE            & 4.620 & 6.483 & 14.541 & 1.779 & 1.502 & 2.509 & 4.275 & 0.388 & 3.995 & 7.005 & 10.809 & 1.070 & 3.717 & 5.938  & 11.043 & 1.371\\
HeteroEmbed    & 5.563 & \underline{7.949} & \underline{17.556} & \underline{2.192} & \underline{3.091} & \underline{5.466} & \underline{7.972} & \underline{0.763} & \underline{5.370} & \underline{9.498} & \underline{13.455} & \underline{1.325} & \underline{6.399} & \underline{10.411} & \underline{17.498} & \underline{1.986} \\
PGPR           & \underline{5.590} & 7.569 & 16.886 & 2.157 & 2.858 & 4.834 & 7.020 & 0.728 & 5.042 & 8.416 & 11.904 & 1.274 & 5.449 & 8.324 & 14.401 & 1.707 \\
\approach\ (Ours)  & \textbf{6.868} & \textbf{9.376} & \textbf{19.692} & \textbf{2.562} & \textbf{3.689} & \textbf{6.340} & \textbf{9.275} & \textbf{0.975} & \textbf{6.313} & \textbf{11.086} & \textbf{15.531} & \textbf{1.692} & \textbf{7.061} & \textbf{10.948} & \textbf{18.099} & \textbf{2.270} \\
\hline
\end{tabular*}
\vspace{-10pt}
\caption{%
Main experimental results evaluating recommendation ranking quality of our method compared to other approaches on four Amazon datasets. The results are computed based on a top-10 recommendation in the test set and are given as percentages (\%). The best results are highlighted in bold font and the best baseline results are underlined.}.
\label{tab:eval}
\vspace{-12pt}
\end{table*}

Next, we introduce the coarse-to-grain process of making explainable recommendations via neural-symbolic reasoning over the knowledge graph.
We start by giving the formal definition of the coarse-grained explanation called \textbf{abstract meta-layout} that provides abstract user behavior patterns.
Formally, an abstract meta-layout $T$ is defined as a rooted tree where each node $x$ is associated with a relation $r_x \in\mathcal{R}$ and a non-negative integer $k_x$ indicating sampling size of entities within node $x$.
As illustrated in Figure \ref{fig:method} (middle panel), the abstract meta-layout can be regarded as a specification of a neural network architecture. We can assemble a tree-structured neural network by instantializing each node $x\in T$ with the corresponding neural relation module $\phi_{r_x}$. 
Each module takes as input the $d$-dimensional vector generated by its parent module and emits a new vector to be passed to the child module.

\vspace{2pt}
\noindent\textbf{Generating Coarse-grained Explanation}\hspace{5pt}
We propose a heuristic approach to generate abstract meta-layout as follows.
Considering that the relation sequence along any root-to-leaf path in the layout forms a metapath, we can construct the layout by simply determining the number of paths to be sampled for each metapath and then combining the set of metapaths that will be used to sample paths. 
Formally, let $y_j$ be the number of paths sampled for metapath $\pi_j$ and $M$ be the total number of metapaths in $\mathcal{G}_\mathrm{R}$.
For any user $u$, we solve the following optimization problem to obtain the optimal $\{y_1,\ldots,y_M\}$:
$\max_{\{y_j\}}~ \sum_j y_j\cdot v(u,\pi_j)~~
\text{subject to}~ \sum_j y_j\le K, 0\le y_j\le k_j,1\le j\le M$,
where $K$ is the maximum number of paths to be sampled in total, $k_j$ is the maximum number of paths sampled for $\pi_j$.
Here, we also introduce a heuristic function
$ v(u,\pi_j) = {\mathbb{E}}_{L\in \mathcal{L}_{u+}^{\pi_j}} \left[\log P(L\mid u,\pi_j)\right] $,
which reflects the estimated importance of metapath $\pi_j$ leading to good items for recommendation to user $u$. 
After obtaining $\{y_1,\ldots,y_M\}$, we can generate an abstract meta-layout $T$ for user $u$ using Algorithm \ref{alg:program-gen}. 
The basic idea is to first construct the layout tree by merging all metapaths in $\{\pi_j \mid y_j>0\}$, which determines the tree structure and the relation in each node.

\vspace{2pt}
\noindent\textbf{Generating Fine-grained Explanation}\hspace{5pt}
Given user $u$ and the abstract meta-layout $T$, the final step is to generate concrete paths in the knowledge graph for recommendation that also serve as detailed explanations.
The path generation process is illustrated in Fig.~\ref{fig:method} (right panel). 
Specifically, as previously discussed, we first assemble a tree-structured neural network based on $T$.
By feeding in the user embedding $\mathbf{u}$, each neural relation module $\phi_{r_x}$ at tree node $x$ can be ``executed'' one-by-one in breadth-first order to output an intermediate $d$-dimensional vector, denoted by $\hat{\mathbf{e}}_{r_x}$.
Then, let $\tau_x$ be the set of paths generated at node $x$ and $\tau_\text{root}=\{\{u\}\}$.
For any other node $x$, we generate the paths as
$\tau_x=\{L\cup\{r_x, e\} \mid L\in \tau_{\text{parent}(x)}, \forall e\in\mathcal{E}~\text{and}~(\text{lastnode}(L), r_x, e)\in\mathcal{G}_R, rank(\langle \hat{\mathbf{e}}_{r_x}, \mathbf{e}\rangle)\le k_x\}$.
The final reasoning paths are collected from all leaf nodes.

\section{Experiments}\label{sec:experiment}

We experiment on four domains of Amazon e-commerce datasets \cite{he2016ups}: \emph{CDs and Vinyl}, \emph{Clothing}, \emph{Cell Phones} and \emph{Beauty}.
Each dataset is considered as an individual benchmark that constitutes a knowledge graph with respectively $388.5M$, $36.5M$, $37.2M$ and $37.3M$ triples. 
We adopt the same training and test splits as \cite{xian2019kgrl}. 
We consider as baselines the latest \emph{in-progress} method PGPR \cite{xian2019kgrl} along with selective baselines included in their study. We also include HeteroEmbed \cite{ai2018learning}, a strong \emph{post-hoc} method for explainable recommendation over KG, which is used as teacher model in Eq. \ref{eq:loss_rank}.
Four metrics are adopted to evaluate all models: Normalized Discounted Cumulative Gain (\textbf{NDCG}), \textbf{Recall}, Hit Rate (\textbf{HR}), and Precision (\textbf{Prec.}).

\begin{figure}[t]
\centering
\begin{minipage}[c]{0.95\linewidth}
	\includegraphics[width=0.325\linewidth,height=0.241\linewidth]{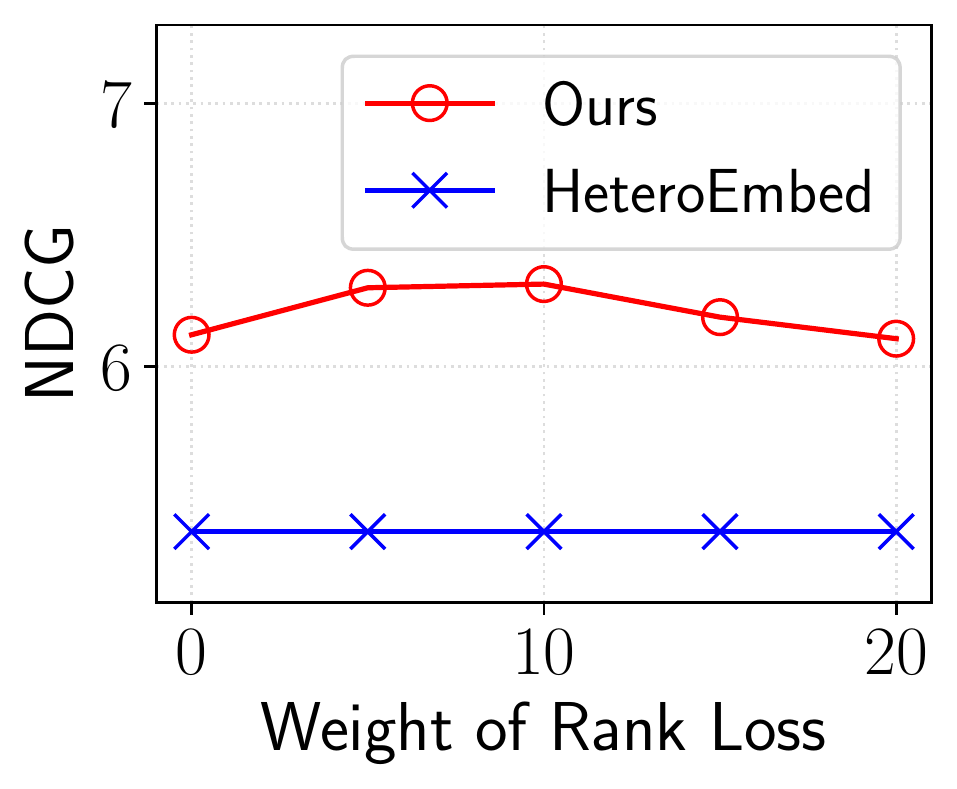}
	\includegraphics[width=0.325\linewidth,height=0.241\linewidth]{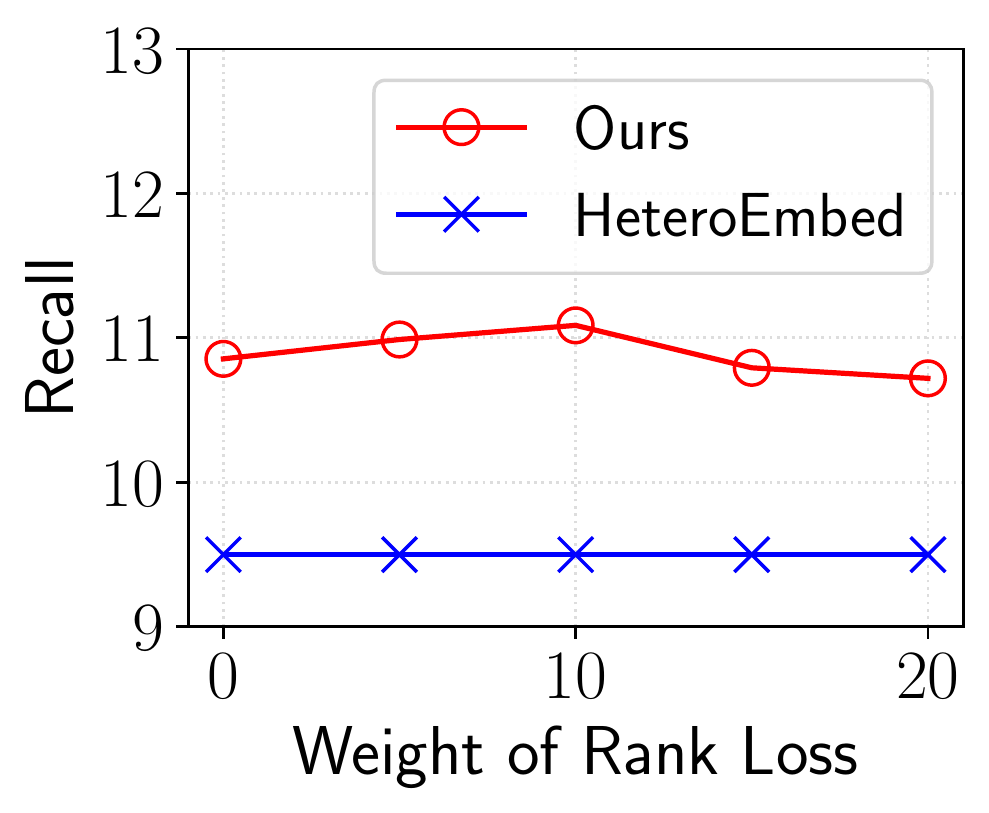}
	\includegraphics[width=0.325\linewidth,height=0.241\linewidth]{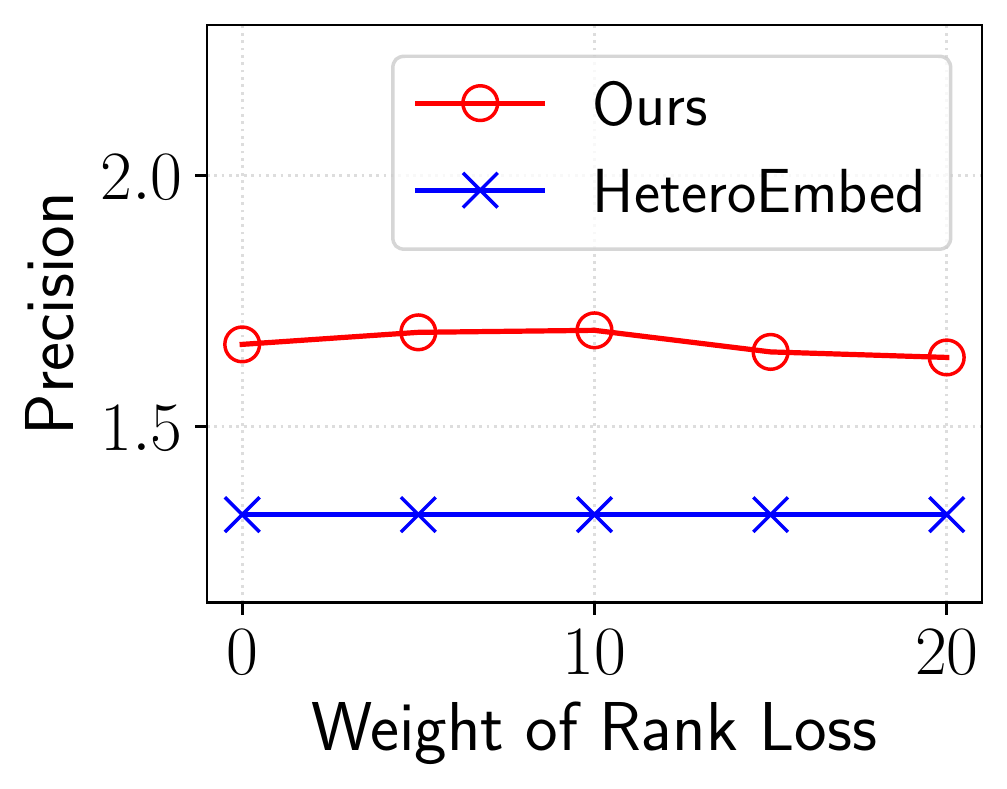} \\
	\vspace{-2pt}
	\hspace*{0.25in} {\footnotesize (a) NDCG \hspace{0.50in} (b) Recall \hspace{0.50in} (c) Precision }
	\vspace*{-5pt}
	\caption{Ranking loss results on Cell Phones dataset.}
	\label{fig:cell_rankweight}
\end{minipage}
\vspace{-5pt}
\end{figure}

\vspace{2pt}
\noindent\textbf{Recommendation Performance}\hspace{5pt}
We first show the recommendation performance of our proposed \approach{} method compared to the baselines.
The results on four datasets are shown in Table \ref{tab:eval}.
We observe that our method outperforms the best baseline HeteroEmbed by a large margin across all settings.
For example, on the Clothing dataset, our model achieves 6.340\% in Recall, which is higher than 5.466\% of HeteroEmbed and 4.834\% of PGPR. Similar trends can also be observed for other metrics.
It is interesting to see our model also outperforms the teacher model HeteroEmbed substantially.
One reason is that the similarity estimation by the teacher network largely reduces the search space for the correct target items, so our model can further discover patterns for each user to make better recommendations.

\vspace{2pt}
\noindent\textbf{Influence of Ranking Loss}\hspace{5pt}
In this experiment, we show how the weighting factor $\lambda$ in Eq.~\ref{eq:loss_all} affects the performance of neural-symbolic representation learning.
We experiment on the Cell Phones dataset and choose different values of $\lambda \in \{0, 5, 10, 15, 20\}$. $\lambda=0$ means no ranking loss is imposed for training. All other settings remain the same as in the previous experiment.
The results are plotted in Fig. \ref{fig:cell_rankweight}, including our model (red curves) and the best baseline HeteroEmbed (blue curves).
We observe two interesting trends.
First, our model consistently outperforms HeteroEmbed under all settings of $\lambda$ in terms of all metrics. Even without the ranking loss, our model can still guarantee a high quality of recommendation. 
Second, a larger weight of the ranking loss may not always entail a better performance. Instead, there is a trade-off between the ranking (Eq.~\ref{eq:loss_rank}) and path regularization (Eq.~\ref{eq:entity2}). The reason is that if the ranking loss plays a dominant role, the model will pay. less attention to the path fitting function, and consequently, it may not be able to find the correct paths to reach a promising item.

\begin{table}[t]
\scriptsize
\begin{tabular}{p{3.2em}|p{2.em}p{2.em}p{2.em}p{2.em}|p{2.em}p{2.em}p{2.em}p{2.em}}
\hline
Dataset & \multicolumn{4}{c|}{\textbf{Cell Phones}} & \multicolumn{4}{c}{\textbf{Beauty}} \\
\hline
Method   & NDCG  & Recall & HR    & Prec.  & NDCG  & Recall & HR    & Prec. \\
\hline
uniform   & 4.545 & 7.229  & 10.192	& 1.087 & 6.293 & 9.256  & 15.564 & 1.918 \\
prior     & 6.255 & 10.842 & 15.097 & 1.659 & 6.880 & 10.393 & 17.258 & 2.224 \\
heuristic & 6.313 & 11.086 & 15.531 & 1.692 & 7.061 & 10.948 & 18.099 & 2.270 \\
\hline
\end{tabular}
\vspace{-10pt}
\caption{Influences of three different abstract meta-layouts.}
\label{tab:program-exp}
\end{table}

\vspace{2pt}
\noindent\textbf{Effectiveness of Abstract Meta-Layout}\hspace{5pt}
Now, we evaluate the effectiveness of our abstract meta-layout for recommendation.
We specifically consider three layouts that are all generated by Algorithm \ref{alg:program-gen} but based on different $\{y_1,\ldots,y_M\}$.
The first one (\emph{uniform}) uses $y_j=\sum_i^K\mathbbm{1}\{X_i=j\}$ where $X_i\in[M]$ is a random variable from a uniform distribution.
The second one (\emph{prior}) uses $y_j$ based on a prior $k_j$, i.e., $y_j=\lceil K\cdot k_j/\sum_ik_i \rceil$. 
The third one (\emph{heuristic}) is our own method, with $y_j$ obtained based on a heuristic function.
The results on Cell Phones and Beauty datasets are reported in Table \ref{tab:program-exp}.
We observe that our generated heuristic program exhibits better recommendation performance than the other two.
This is because the heuristic value $v(u,\pi_j)$ that measures the quality of metapath $\pi_j$ is directly estimated from neural symbols.

\section{Conclusions and Future Work} \label{sec:conclusions}
In this paper, we study the problem of explainable recommendation by explicitly reasoning over knowledge graph. 
We propose a neural-symbolic approach (\approach{}) that absorbs both explainability of symbolic reasoning and generalization ability of deep learning.
The model generates two-stage coarse-to-fine explanations, where coarse-grained explanation describes representative user pattern and is used to generate fine-grained explanation for final explanation and recommendation.
We extensively evaluate our model on several real-world benchmarks and show that the approach delivers strong recommendation results.

\bibliography{paper}
\bibliographystyle{aaai}
\end{document}